# The Mechanical Yes-Man:
# Emancipatory AI Pedagogy in Higher Education

**Abstract:** The proliferation of Large Language Models in higher education presents a fundamental challenge to traditional pedagogical frameworks. Drawing on Jacques Rancière's theory of intellectual emancipation, this paper examines how generative AI risks becoming a "mechanical yes-man" that reinforces passivity rather than fostering intellectual autonomy. Generative AI's statistical logic and lack of causal reasoning, combined with frictionless information access, threatens to hollow out cognitive processes essential for genuine learning. This creates a critical paradox: while generative AI systems are trained for complex reasoning, students increasingly use them to bypass the intellectual work that builds such capabilities. The paper critiques both techno-optimistic and restrictive approaches to generative AI in education, proposing instead an emancipatory pedagogy grounded in verification, mastery, and co-inquiry. This framework positions generative AI as material for intellectual work rather than a substitute for it, emphasising the cultivation of metacognitive awareness and critical interrogation of AI outputs. It requires educators to engage directly with these tools to guide students toward critical AI literacy, transforming pedagogical authority from explication to critical interloping that models intellectual courage and collaborative inquiry.

## 1. A Wake-Up Call

The proliferation of Large Language Models in Higher Education is a wake-up call many seasoned academics would rather ignore. Some are still sleeping through it, turning their backs and hoping it will stop. Others have jolted awake in a panic, attempting to restrict usage without understanding what they are really up against.

The call deep-down whispers: *the world you grew up in no longer exists.*

The natural response – that 'things were better before' – while capturing legitimate anxieties, becomes a paralysing fallacy that impedes the critical engagement this moment demands. This tendency is particularly seductive in academic contexts, where established methodologies carry institutional weight and technological change challenges fundamental pedagogical assumptions.

In his radical theory of education, first published in French in 1987, Jacques Rancière distinguishes between genuine intellectual progress and its institutional counterfeit. He describes the man of progress as 'a man who moves forward, who goes to see, experiments, changes his practice, verifies his knowledge, and so on without end [...] not concerned with the social rank of someone who has affirmed such and such a thing, but [who will] go see for himself if the thing is true.'[1] This figure stands in striking contrast to what Rancière calls the "opinion of progress", which assumes a point of departure already inscribed within established hierarchies: it takes for granted that some know more, and that others must catch up by following their path.

Whereas the man of progress verifies and experiments without regard to rank, the opinion of progress entrenches the explicative order, defining advancement as moving along a track laid down by those presumed more enlightened. Explication, Rancière insists, is the fiction of pedagogy and the weapon of pedagogues. It creates a structure of delay and an imaginary distance between knowing minds and ignorant ones, between the capable and incapable, thereby reinforcing the very bond of the social order and leading to enforced stultification. This logic subordinates one intelligence to another, creating a closed circle of power that sustains the hierarchy of the superior and the inferior.

---

[1] Jacques Rancière, *The Ignorant Schoolmaster: Five Lessons in Intellectual Emancipation*, trans. Kristin Ross (Stanford, CA: Stanford University Press, 1991), 109.

In current debates on generative AI use in higher education, polarised opinions of progress shift between a conservative, new-Luddite framing that views the technology as a threat to established hierarchies of knowledge and a techno-optimistic excitement for anything new, as if that alone were progress. Across this spectrum, while European institutions are taking a more cautious regulatory approach, universities in China, Singapore, and UAE are adopting a markedly different stance. Nearly 60% of Chinese university faculty and students report using AI tools frequently[2], while in Singapore 86% of university students now use AI tools in their studies,[3] and UAE students are among the most frequent users globally with 32% using AI tools weekly.[4] Institutions in these countries treat AI literacy as a skill to be mastered rather than a problem to be managed.

Despite these different approaches, a fundamental risk remains: that generative AI in education may hollow out the very capacities it claims to cultivate. Recent studies show [5] [6] that overreliance on AI in learning environments reduces critical thinking skills and cognitive engagement as students increasingly offload their thinking processes. In Rancière's terms, generative AI risks assuming the role of the explainer – an all-knowing, mechanical yes-man that positions itself as the source of understanding, reinforcing passivity and dependency rather than fostering intellectual autonomy.

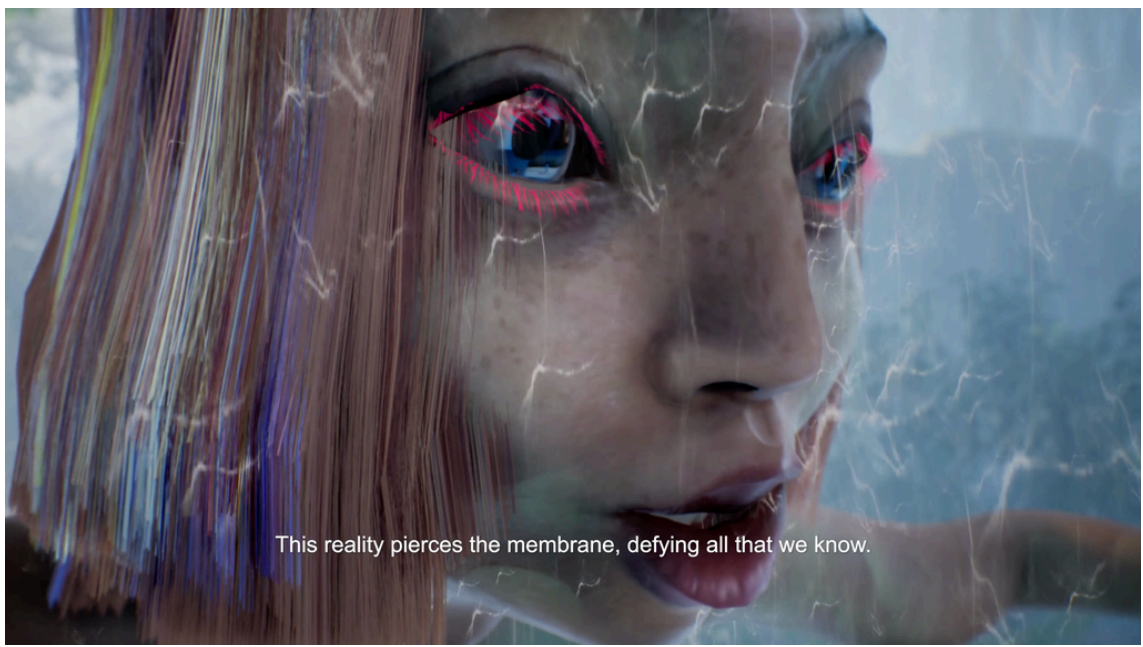

Figure 1: Keiken co-led by Tanya Cruz + Hana Omori made in collaboration with sound by Flore (Khidja), *Dream Time Life Simulation*, courtesy of the artist. © Keiken, 2021.

---

[2] Caiwei Chen, "Chinese Universities Want Students to Use More AI, Not Less," *MIT Technology Review*, July 28, 2025, https://www.technologyreview.com/2025/07/28/1120747/chinese-universities-ai-use/.
[3] Digital Education Council. "Digital Education Council Global AI Student Survey 2024." August 2, 2024. https://www.digitaleducationcouncil.com/post/digital-education-council-global-ai-student-survey-2024.
[4] "Anthology Survey Reveals Enthusiastic Approval of AI in UAE Universities," Anthology, November 2, 2023, https://www.anthology.com/news/anthology-survey-reveals-enthusiastic-approval-of-ai-in-uae-universities.
[5] Michael Gerlich, "AI Tools in Society: Impacts on Cognitive Offloading and the Future of Critical Thinking," *Societies* 15, no. 1 (2025): 6, https://doi.org/10.3390/soc15010006.
[6] Nataliya Kosmyna et al., "Your Brain on ChatGPT: Accumulation of Cognitive Debt when Using an AI Assistant for Essay Writing Task," *arXiv* (2025), https://arxiv.org/abs/2506.08872.

## 2. The Crisis of Learning

Generative AI's tendency to perform as a mechanical yes-man is reinforced by the way it sidesteps causality, producing convincing answers without the underlying reasoning that meaningful learning demands. Operating primarily on statistical logic, generative AI systems excel at finding correlations but they don't grasp the "why" behind these relations. Without causal grounding, explanations feel effortless to obtain but lack the logical scaffolding that supports genuine learning and, therefore, risk being accepted without verification, even when they are fabricated or false. This ease is precisely the danger to the learning process: it bypasses the intellectual labour through which learners develop the circuits of reasoning, critical analysis, and independent judgement.

The essence of learning lies in engaging with intellectual effort and the time it requires – yet AI's promise of immediate solutions and frictionless information access collapses the reflective distance necessary for deep thinking and understanding. The erosion of this temporal dimension needed for the thinking process to mature undermines the foundations of genuine inquiry while reducing the motivation to seek more profound understandings.

This paradigm creates a profound irony. While developers train AI systems for human-like, complex reasoning, students increasingly use these tools to deskill themselves and bypass the very intellectual work that builds such high-level capabilities. What emerges is a broader crisis of learning: as our tools become more intelligent, our engagement with intellectual work risks becoming more passive. The very technologies designed to extend human intelligence may be undermining the cognitive processes through which learning, seen as a transformative engagement with the world, becomes growth.

If, as Rancière argues, the primary goal of education is emancipation, 'to take the measure of someone's intellectual capacity, and decide how to use it'[7], then what pedagogical approaches and skill development can lead to emancipation in a higher education landscape where multiple forms of intelligence and ways of learning overlap amid intense job market competition?

The challenge here is not really whether AI should be used in higher education. This is not ours to settle; AI is already transforming how students think, write, and learn, and attempts to police it are both impractical and futile. The real task is to determine how its integration into teaching and learning can resist becoming another vector for stultification, reinforcing hierarchies of knowledge under the guise of innovation, and instead be engaged with in ways that expand critical and creative abilities.

All too frequently, the purpose of higher education has been narrowed to producing outputs for the knowledge economy, with success measured in metrics, efficiency targets, and rankings. This builds on decades of neoliberal policy that has marketised education, particularly influential reforms from England with global reach, leading to deprofessionalised teaching and stagnant learning outcomes.[8] AI accelerates this drift, shaking to its core the very motivations for learning. What is education for? When students pay substantial sums for their higher education, what are they purchasing beyond credentials? This narrowing reflects what Gert Biesta calls 'learnification'[9] – the reduction of education to learning, where the focus shifts from educational values to the mere acquisition of knowledge, skills, and competencies. Learnification transforms students into consumers of learning experiences while obscuring fundamental questions about what should be learned, why it matters, and what kind of professional one is trying to become. By offering seemingly frictionless access to information and immediate problem-solving, AI creates the illusion of educational progress while bypassing the formative encounters with difficulty, uncertainty, and otherness that constitute genuine educational experience.

---

[7] Rancière, *The Ignorant Schoolmaster*, 17.
[8] Stephen J. Ball, *The Education Debate: Policy and Politics in the Twenty-First Century*, 4th ed. (Bristol: Policy Press, 2017).
[9] Gert J. J. Biesta, *Good Education in an Age of Measurement: Ethics, Politics, Democracy* (Boulder, CO: Paradigm Publishers, 2010).

These answers are complicated by the digital infrastructure in which education now operates. As Shoshana Zuboff observed,[10] digital architectures offer new opportunities for participation while tightening measures of control and commodification. This pattern has intensified as surveillance capitalism has progressively evolved into what Yanis Varoufakis[11] argues represents a shift toward technofeudalism: a system where platforms function not merely as market participants but as digital lords of the cloud, controlling both infrastructure and participation while extracting rent from the digital footprints of all who engage within their domains.

For educational AI tools, this creates a subtle but significant tradeoff: the same AI tools that offer efficiency, assistance, personalised learning, and immediate feedback also introduce patterns of behavioural monitoring, data and labour extraction, and platform dependency with profound, albeit externalised, social and environmental consequences. More fundamentally, this signals a transfer of knowledge infrastructures from public institutions to private commercial platforms, where knowledge once accessible through public systems becomes mediated by proprietary AI services. This privatisation simultaneously reshapes how knowledge is constructed, as the filtering of information through algorithmic logics leads to a progressive epistemological reductionism. In turn, this invisible structuring of access and engagement alters personal agency and autonomy.

3. **Verification and Mastery**

Rancière's core provocation – that we must begin from equality of intelligence and sustain it through verification – is especially urgent. For Rancière, emancipation is not a destination but a practice of self-verification and inquiry, where will and attention drive genuine learning rather than passive consumption of explanations. In the context of AI-assisted learning and education, this framework becomes particularly vital. Given generative AI's limitations in causal reasoning, the act of verification transforms from a pedagogical ideal into a practical necessity. Students must learn to interrogate AI outputs, cross-reference claims, and engage in the kind of critical research that AI cannot perform. Rather than accepting AI as the new explainer – the mechanical source of ready-made understanding – emancipatory education requires using AI as material for intellectual work, not a substitute for it. The student's will and attention become even more crucial when the temptation to outsource thinking is so readily available.

These pedagogical challenges are compounded by the acceleration of information flows and the fractured temporalities shaped by smartphones and social media. As Walter Benjamin cautioned[12], the critique of progress must address not only its goals but its pacing and progression. Paul Virilio later theorised[13] speed as the primary force driving modern civilisation, arguing that acceleration compresses time for reflection and decision-making. This velocity imperative drives the design of digital platforms, as algorithmically optimised feeds cultivate habits of rapid scanning and surface engagement. Applied to educational contexts, such acceleration can foreclose reflection, obscure alternatives, and narrow the space for verification. Years of interaction with these systems have shortened attention spans, normalised cognitive fragmentation, and made the deliberate allocation of time to sustained inquiry an increasingly rare discipline.

---

[10] Shoshana Zuboff, *The Age of Surveillance Capitalism: The Fight for a Human Future at the New Frontier of Power* (New York: PublicAffairs, 2019).
[11] Yanis Varoufakis, *Technofeudalism: What Killed Capitalism* (London: Bodley Head, 2023).
[12] Walter Benjamin, "Theses on the Philosophy of History" (1940), in *Illuminations*, ed. Hannah Arendt, trans. Harry Zohn (New York: Schocken Books, 1968), 253-264.
[13] Paul Virilio, *Speed and Politics*, trans. Mark Polizzotti (New York: Semiotext(e), 1986).

Against this backdrop, AI presents both the greatest threat to intellectual emancipation and, paradoxically, a particularly urgent catalyst.

The need for critical engagement gains renewed momentum when we consider how learning has already been transformed by networked media. For years, students have turned to Reddit communities, YouTube tutorials, online forums and podcasts, often finding greater engagement there than in traditional classrooms. This distributed media landscape has fundamentally blurred Rancière's division between the 'producers' and 'consumers' of knowledge – a shift Alvin Toffler anticipated in *The Third Wave* (1980), where he predicted that in the post-industrial age, the roles of producers and consumers would merge into the "prosumer." This describes how we all engage with networked, and now neural media: consuming content while simultaneously producing responses, analyses, and insights through comments, posts, and algorithmic interactions, as a form of situated knowledge production. This phenomenon demonstrates that intellectual equality is not just a philosophical ideal but an active practice in many corners of the internet, often in 'dark forests'[14] that thrive outside the most commercial platforms. There, knowledge is understood as contingent and relational.

Generative AI now enters this landscape with the potential to either deepen these practices of intellectual equality or erode them, reinstating the hierarchy of the explainer and reducing active participants to passive recipients. This represents both an extension and intensification of existing trends: just as students learned to navigate the vast, unverified information landscape of the internet – developing skills to distinguish reliable sources from misinformation and to synthesise across perspectives – they must now cultivate parallel competencies with AI. The difference is that AI's sophistication makes the stakes higher and the need for critical engagement more urgent.

Beyond techno-optimistic speculation, the economic reality of AI literacy as professional currency cannot be ignored and creates opportunities for much-needed arts and humanities engagement. Globally, approximately 86% of students now use AI tools regularly in their learning,[15] making avoidance increasingly impractical. In the UK, despite questions about whether current market enthusiasm reflects sustainable trends or temporary hype, the demand for AI-related skills has grown, with job postings rising by 21%, and AI competencies commanding a 23% wage premium, exceeding that of formal degrees up to PhD level.[16] In this context, true autonomy and emancipation require not rejection but mastery – the ability to navigate AI tools critically and creatively rather than being used by them. This means understanding their potential and limitations, interrogating their outputs and usage, and maintaining the critical autonomy that prevents technological dependence from becoming intellectual subordination.

This need for mastery extends to educators too. While the vertical master-schooling paradigm and pedagogical authority that Rancière critiqued was already showing cracks before his intervention, AI now threatens the very possibility of universal teaching that Jacotot championed. We cannot, in fact, follow Jacotot's example of teaching a language without knowing it when it comes to AI. The technology needs demystification and critical intervention; it cannot be approached from a position of wilful ignorance.

---

[14] Yancey Strickler, "The Dark Forest Theory of the Internet," *The Ideaspace*, May 2019, https://www.ystrickler.com/the-dark-forest-theory-of-the-internet/
[15] Kelly, Rhea. "Survey: 86% of Students Already Use AI in Their Studies," *Campus Technology*, August 28, 2024, https://campustechnology.com/articles/2024/08/28/survey-86-of-students-already-use-ai-in-their-studies.aspx
[16] Eugenia Gonzalez Ehlinger and Fabian Stephany, "Skills or Degree? The Rise of Skill-Based Hiring for AI and Green Jobs," *SSRN Electronic Journal*, February 25, 2024. https://doi.org/10.2139/ssrn.4603764

Unlike Jacotot, who could guide students through Telemachus without speaking Dutch himself, educators today cannot effectively guide students through critical AI literacy without engaging directly with these tools. They must test AI systems, probe their limitations, understand their biases, and experience both their capabilities and failures firsthand. This is not a betrayal of Rancière's principles but their practical application: educators need not be AI experts, but they must be explorers – approaching these tools with the same spirit of inquiry and verification they hope to cultivate in students. Understanding AI's capabilities and blind spots becomes essential for this exploratory practice – knowing where these tools excel and where they inevitably fail, while developing critical awareness of their ontological, epistemological, political, economic, and environmental implications.

More broadly, it is urgent that researchers and practitioners from the arts and humanities engage directly with AI development, bringing the critical perspectives necessary to broaden their foundational assumptions, rather than leaving their design entirely to computer scientists and engineers. As AI continuously redistributes information and intellectual legitimacy, and students become more globally connected and digitally literate, the role of the educator must be reconfigured.

If educators retreat from AI out of fear or principle, they abdicate their role in helping students develop critical relationships with these tools. This institutional delay paradoxically recreates Rancière's structure of delay – not the artificial postponement of the explicative order, but a temporal gap that creates new hierarchies of 'velocity' where some navigate AI's complexities alone, potentially reinforcing the very dependency and intellectual subordination that emancipatory education seeks to prevent.

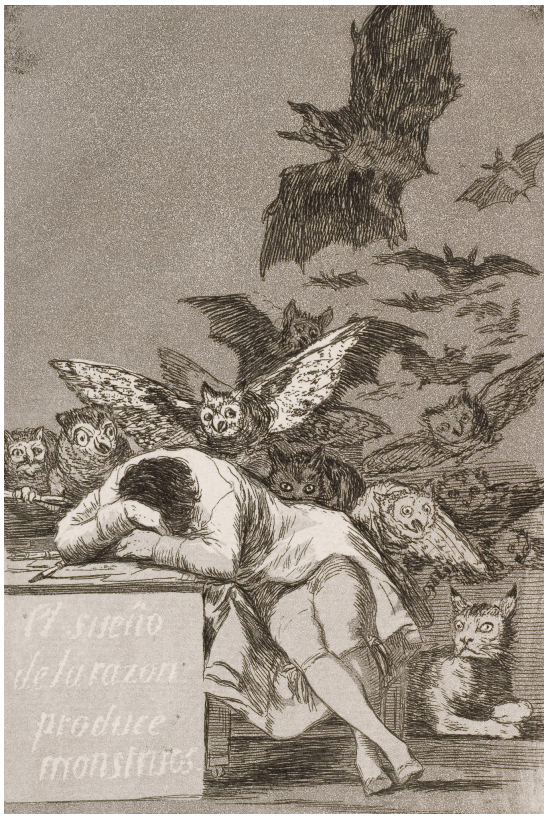

Figure 2: Francisco Goya, *The Sleep of Reason Produces Monsters*, circa 1799. Etching from Los Caprichos (no. 43).

## 4. Emancipatory Pedagogy in Practice

In this new ecosystem, where knowledge flows across students, educators, and forms of intelligence, authority is distributed. The monopoly of meaning-making has been broken wide open and requires collaborative efforts to rewire some fundamental truths around learning.

If higher education is to respond meaningfully to these transformed conditions, it must adopt new pedagogical methods. Not a method founded on superior knowledge but on co-inquiry, verification, and intellectual quest. Instead of explicative approaches, emancipatory pedagogy requires a method that emerges through negotiation and interaction – one that develops intellectual capacities that can't be easily automated: the ability to engage in sustained inquiry, to collaborate with human and other intelligences, and to adapt thinking as circumstances change. This is what students' investments secure: access to communities of practice where they engage with others wrestling with similar intellectual challenges. Rather than functioning as individual consumers of knowledge, they become members of a shared social and democratic space where collective agency emerges through negotiated inquiry, supported by the time and guidance necessary for scaffolding intellectual development and critical independence. This approach honors Rancière's vision of 'the mind as activity,' where 'intelligence is attention and research before being a combination of ideas.'[17]

This requires creating spaces for co-learning with students while developing patterns that foster critical and productive engagement. Educators must cultivate students' metacognitive and ethical awareness of how these systems operate, providing the critical frameworks that enable the questioning of sources, assumptions, and implications of AI-generated content. The authority of the educator transforms into that of a critical interloper – someone who intervenes in the AI-student interaction to inject critical questioning, modeling intellectual humility while empowering learners to discern, question and deconstruct machine-generated claims.

The contemporary educator faces a dual imperative: to equip students with the capacity to critically interrogate AI outputs and to model ethical and reflexive engagement with technology in their own practice. The goal is not to teach students what to think about AI, but to teach them to think independently – to learn by doing and, if desired, use these tools to expand rather than replace their thinking. This means bringing knowledge acquisition closer to knowledge application, ensuring that students encounter and work with knowledge in contexts where they must verify, question, and apply it, making personal connections that give that knowledge meaning.

Ultimately, this approach emphasises collaboration over control, curiosity over compliance. It encourages students to explore their interests, to research meticulously, to follow their intellectual drives and sharpen their thinking– using AI as a tool for discovery rather than a crutch for avoidance. Such practices cultivate the intellectual resilience needed for the technosocial shifts that will inevitably come their way.

The world we, as educators, grew up in no longer exists, nor do the assumptions that once underpinned our learning. In this transformed landscape, we bear the responsibility of modeling the very intellectual courage and critical engagement we seek to cultivate in our students. This moment demands rewiring education around the urgent challenge of practising agency and pursuing emancipation in an age where knowledge abounds and intelligence is contested.

This article was written with assistance from Claude Sonnet 4.

---

[17] Rancière, *The Ignorant Schoolmaster*, 54.